# Quantization as a Necessary Condition for Gauge Invariance

J. Towe

Department of Physics, The Antelope Valley College
Lancaster, CA, USA
jtowe@avc.edu

Symmetry and quantization are the two major enterprises of theoretical physics; but some argue that quantization can be derived as a necessary condition for symmetry. It is argued here that the Heisenberg uncertainty principle is a necessary condition for gauge invariance in the 5-dimensional Kaluza-Klein theory.

## 1. Introduction

Quantum mechanics has superceded all classical theories other than the general theory of relativity (GTR), and indeed, it has traditionally been expected that GTR will ultimately be regarded as an approximation of a quantum theory of gravity. Einstein rejected this view however, and argued that quantum mechanics would ultimately give way to classical physics. It was traditionally a 'hard call', because both GTR and quantum mechanics are incomplete. While GTR cannot account for stellar collapse beyond the singularities that characterize the theory, quantum mechanics cannot account for individual events. But classical theory contains more information than quantum theory, and in this context, it was argued that a classical theory can become a quantum theory if the former, for some reason, loses information . In this view, classical theory is regarded as a union of quantum mechanics and certain 'hidden variables'. The approach involving hidden variables was considered by Einstein, but ultimately discarded by him, and by the physics community, because hidden variable theories produced no new physical results. Because the hidden variable approach seems fundamentally sound however, it has been revived. One such theory was proposed by this author. It was demonstrated that the Bohr radii constitute a necessary condition for gauge invariance in the 5-dimensional Kaluza-Klein theory. Specifically, it was demonstrated that if GTR on 5-spacetime is projected onto 4-spacetime, so that one spacetime dimension becomes a hidden variable, and if GL(5) reduces to GL(4)XU(1), then the Bohr radii constitute a necessary condition for U(1) invariance on the micro-scale [J. Towe. 1988]. A second such theory was proposed by B. Muller, who demonstrated that a classical system in five dimensions becomes a quantum system if observed in four dimensions [B. Muller, 2001]. A complementary result is due to R. Bousso, who derived the Heisenberg relation from the holographic limit [R. Bousso, 2004]. Complimenting these results, G. 't Hooft contributed an argument, that classical theories may lose information due to dissipative forces [G. 't Hooft, 1999].

The conflict between general relativity and quantum mechanics and the debate, which assumes that one theory must displace the other as the logical bottom line could be avoided if GTR and quantum mechanics could be unified into a new theory that would become general relativity in the large scale domain and quantum mechanics in the small. An early attempt to achieve this ideal was introduced in the context of Hermann Weyl's theory of gravitation and classical electrodynamics [H. Weyl, 1918]. This theory applied a generalization of Riemannian geometry in which vectors can undergo increments both of orientation and scale when parallel transported around a closed curve. The degrees of freedom that result from this hypothesis are precisely sufficient to produce Einstein's gravitation in the astrophysical domain of 4-spacetime and classical electrodynamics in the macroscopic domain. The discovery that the Weyl theory also requires quantization conditions occurred when Einstein critiqued Weyl's theory, observing that the orbital motions of electrons would, in this theory, associate with increments of proper time scale that would contradict observation [H. Weyl, 1922]. This claim motivated F. London to demonstrate that the Klein-Gordon equation can be derived, in the context of the Weyl theory, as a necessary condition for the preservation of proper time under Weyl scale transformations [F. London, 1927]. Although these results are interesting, the theories of

Weyl and London were not extensively studied, because they produced no new physical results.

The theory that is introduced here parallels the theory of London, and constitutes a generalization of the theory which was introduced by this author in 1988. The fifth spacetime dimension is again regarded as a hidden variable and discarded; i.e. general relativity in five dimensions is again projected onto 4-spacetime; and it is required that the symmetry reduction GL(5) → GL(4)XU(1), which is associated with the projection, be respected by interactions on the projected domain. The relevance of the proposed generalization of the 1988 theory is that it derives general quantization (the Heisenberg relations) as a necessary condition for gauge invariance. The proposed theory also parallels the theory of Muller in that it produces a quantum mechanical version of a classical system by hiding a fifth spacetime dimension, and complements the theory of Bousso in that it establishes the very general Heisenberg relation. The proposed theory should also be compared with the thinking of van de Bruck, who speculated that the dissipative forces, discussed by 't Hooft, may be related to gravity [C. van de Bruck 2000]. But the approach proposed here is strictly in the tradition of Einstein, because it is exclusively differential geometric in character, and derives general quantization conditions from the 5-dimensional theory of general relativity. Finally, it should be observed that the theory proposed here appears to be complemented by recent discussions regarding an ultimate superunification. Specifically, the proposed theory models, from a strictly differential geometric perspective, the (quantum electrodynamical) stationary states of Dirac and Klein-Gordon particles; e.g. of electrons. Based upon recent results, it appears that supergravity in the context of superstring compactification, can be regarded as modeling (supergravitational) stationary states of the baryon [J. Towe, 2005].

## 2. Gauge Invariance in the 5-dimensional Kaluza-Klein Theory

Let us recall the Christoffel connections (of the second kind) that underlie general relativity mechanics on 4-spacetiime:

$$\begin{Bmatrix} \mu \\ \nu \quad \rho \end{Bmatrix} = \frac{g^{\mu\alpha}}{2} \left( \frac{\partial g_{\alpha\nu}}{\partial x^{\rho}} + \frac{\partial g_{\rho\alpha}}{\partial x^{\nu}} - \frac{\partial g_{\nu\rho}}{\partial x^{\alpha}} \right): \quad (1)$$

μ, ν, ρ = 1, 2, 3, 4 [A. Einstein, 1915]. An analogous set of connection coefficients characterize the general theory of relativity on 5-spacetime that was considered by Theodor Kaluza and Oscar Klein

$$\begin{Bmatrix} M \\ N \quad R \end{Bmatrix} = \frac{g^{MA}}{2} \left( \frac{\partial g_{AN}}{\partial x^{R}} + \frac{\partial g_{RA}}{\partial x^{N}} - \frac{\partial g_{NR}}{\partial x^{A}} \right): \quad (2)$$

M, N, R = 1, 2, 3, 4, 5 [T. Kaluza, 1921; O. Klein, 1926; O. Klein, 1938]:

. The coefficients

$$\begin{Bmatrix} \mu \\ \nu\ 5 \end{Bmatrix} = \frac{g^{\mu\alpha}}{2}\left(\frac{\partial g_{\alpha\nu}}{\partial x^5} + \frac{\partial g_{5\alpha}}{\partial x^\nu} - \frac{\partial g_{\nu 5}}{\partial x^\alpha}\right): \quad (3)$$

μ, ν = 1, 2, 3, 4 clearly describe the connection coefficients that are, in addition to those described by (1), perceived on 4-spacetime in the context of the Kaluza-Klein theory. Moreover, the 'Kaluza-Klein' condition

$$\frac{\partial g_{\alpha\nu}}{\partial x^5} = 0 \quad (4)$$

reduces the coefficients (3) to

$$F^\mu{}_\nu = \frac{g^{\mu\alpha}}{2}\left(\frac{\partial g_{5\alpha}}{\partial x^\nu} - \frac{\partial g_{\nu 5}}{\partial x^\alpha}\right), \quad (5)$$

which Kaluza and Klein interpreted as the electromagnetic field tensor on 4-spacetime. Let us now consider the energy, $\mathcal{L}$, of this electromagnetic field in the presence of 4-current. If the energy, $\mathcal{L}$, is described in terms of $F^\mu{}_\nu$, if $\mathcal{L}$ is invariant under the group U(1) of 1-parameter phase transformations (to be discussed), and if the variational principle

$$\delta\int \mathcal{L}\, d^4x = 0,$$

is restricted to a variation of action with respect to the metrical coefficients $g_{\mu 5}$, then the Euler-Lagrange equations are the Maxwell equations (that describe classical electromagnetic interactions in 4-spacetime). If these are integrated, the result is a wave equation, which describrs the propagation of a classical electromagnetic wave in 4-spacetime.

A second postulated result of the projection that is required by (4) is the reduction

$$GL(5) \rightarrow GL(4) \times U(1),$$

where GL(5) and GL(4) respectively represent the groups of general rotations (to be defined) on 5-spacetime and 4-spacetime, under which the equations of the 5-dimensional theory of general relativity and the 4-dimensional theory of general relativity are invariant; and where U(1) represents the above described group of 1-parameter phase transformations.

The sets GL(N) and U(1) of transformations are realizations of 'groups' because the rules that are satisfied by the multiplicative combinations of the exponential transformations include

1. associativity

and
2. commutativity (if abelian group),

and the elements themselves include

3. a multiplicative identity
and
4. a multiplicative inverse.

The elements $\exp(\sum_{\alpha=1}^{N(N-1)/2} G_\alpha u^\alpha)$ that constitute the general linear group GL(N) can be expanded about the identity to yield

$$I + \sum_{\alpha=1}^{N(N-1)/2} G_\alpha u^\alpha + \ldots,$$

where $u^\alpha$ and $G_\alpha$ respectively represent the parameters and generators of the GL(N) group. To first order in the parameters $u^\alpha$, the above expansion reduces to

$$I + \sum_{\alpha=1}^{N(N-1)/2} G_\alpha u^\alpha.$$

GL(N) is said to be a continuous group because the terms $\sum_{\alpha=1}^{N(N-1)/2} G_\alpha u^\alpha$ can be infinitesimal (so that expansion about the identity can be continuous).

The number of parameters N(N-1)/2 represents the difference between the number of variables to be determined and the number of degrees of freedom that are involved in the transformation equations (for flat 4-spacetime; i.e. Minkowski spacetime, these transformation equations are the Lorentz transformations). Thus, N(N-1)/2 is the number of parameters that is required to uniquely determine the group GL(N). The group GL(4) acts upon 4-momentum, so that, by Noether's theorem, the invariance of an interaction under GL(4) is equivalent to the conservation, by that interaction, of 4-momentum.

The group U(1) consists (to first order in the parameter $u = u_\mu dx^\mu$) of the infinitesimal phase transformations

$$\exp \sum_{\mu=1}^{4} u_\mu dx^\mu I = I + \sum_{\mu=1}^{4} u_\mu dx^\mu. \tag{6}$$

U(1) is parameterized in terms of electrical charge, so that invariance of an interaction under U(1) is, by Noether's theorem, equivalent to conservation by that interaction of electrical charge. In the Kaluza-Klein theory however, the tensorial equations describing the acceleration of an electrical charge in an electromagnetic field are

$$m\frac{d^2x^\mu}{d\tau^2} = \begin{Bmatrix} \mu \\ \nu\ 5 \end{Bmatrix} e\left(\frac{dx^\nu}{d\tau}\right) = eF^\mu{}_\nu \frac{dx^\nu}{d\tau} : \mu = 1, 2, 3, 4, \qquad (7A)$$

where e is the electrical charge on the test particle, where m is the mass of the test particle, where $m\frac{d^2x^\mu}{d\tau^2} = eF^\mu{}_\nu \frac{dx^\nu}{d\tau} : \mu = 1, 2, 3, 4$ describes the Lorentz force and where we have adopted the Einstein summation convention (summing over repeated Greek indices from 1 through 4, and over repeated Latin indices from 1 through 3). Thus, comparing equation 7A with the equations of motion of a test particle in a gravitational field:

$$m\frac{d^2x^\mu}{d\tau^2} = \begin{Bmatrix} \mu \\ \nu\ \rho \end{Bmatrix} m\left(\frac{dx^\nu}{d\tau}\right)\frac{dx^\rho}{d\tau} = \begin{Bmatrix} \mu \\ \nu\ \rho \end{Bmatrix} P^\nu \frac{dx^\rho}{d\tau} : \mu = 1, 2, 3, 4, \quad (7B)$$

one observes that, in the Kaluza-Klein theory, electrical charge, e, constitutes the 5[th] component of 5-momentum as perceived in 4-spacetime, so that 5-action is given by

$$edx^5 - P_\mu dx^\mu.$$

But, just as 4-action $Edt - P_j dx^j$ reduces to 3-action $Edt = P_j dx^j$ on 3-spacetime, 5-action $edx^5 - p_\mu dx^\mu$ reduces to 4-action $edx^5 = p_\mu dx^\mu$ on 4-spacetime. In the 4-spacetime projection of GTR, which is proposed by the Kaluza-Klein theory then, parameterization in terms of electrical charge $edx^5$ is equivalent to parameterization in terms of 4-momentum $P_\mu dx^\mu$.

In this context, consider a classical wave propagating in 4-spacetime; and consider the transformation of this wave under U(1):

$$\hat{\Psi}^\alpha = \Psi^\alpha \exp i \frac{P_\mu dx^\mu}{k}, \qquad (7)$$

where k is the smallest positive value with the dimensions of action. The invariance of this wave state under the group U(1) clearly implies that

$$\hat{\Psi}^\alpha = \Psi^\alpha \exp i \frac{P_\mu dx^\mu}{k} = \Psi^\alpha \exp i(2\pi n) = \Psi^\alpha \cos(2\pi n) = \Psi^\alpha, \qquad (8)$$

where $P_4 = -E/ic$, $dx^4 = icdt$ and where n is an integer; or that

$$\frac{P_\mu dx^\mu}{k} = 2\pi n. \qquad (9)$$

Since division by zero is mathematically precluded, k is non-zero. Moreover, because negative values of action, $P_\mu dx^\mu$, are not considered, n is positive. It is therefore implied that

$$P_\mu dx^\mu \geq 2\pi k. \qquad (10)$$

Moreover, if axes are aligned so that motion is parallel to the μ-axis, one can write

$$P_\mu dx^\mu \geq 2\pi k, \qquad (11)$$

where summation is not implied. Thus, since every axis can be aligned with motion, one can write

$$P_\mu dx^\mu \geq 2\pi k: \mu = 1, 2, 3, 4. \qquad (12)$$

Finally, since k has the dimensions of action, one can assign to k a constant value:

$$k = \frac{\hbar}{4\pi}, \qquad (13)$$

so that (11) reduces to

$$P_\mu dx^\mu \geq \frac{\hbar}{2}: \mu = 1, 2, 3, 4; \qquad (14)$$

which are the Heisenberg relations on 4-spacetime [L. Schiff, 1968].

### Conclusion

The 5-dimensional Kaluza-Klein theory was revisited, and it was demonstrated that if one hides the variable that is represented by the fifth spacetime dimension (if one projects the 5-dimensional theory of general relativity onto 4-spacetime; and if interactions on the projected domain are constrained to respect the symmetry that is imposed upon 4-spacetime by the postulated projection (the symmetry reduction GL(5) → GL(4)XU(1)), then general relativity on 5-spacetime projects onto the astrophysical domain of 4-spacetime as general relativity; onto the macroscopic domain of 4-spacetime as classical electrodynamics and onto the microscopic domain of 4-spacetime as quantum mechanics.
    The notion that quantum mechanics can emerge from classical theory; e.g. from general relativity is often described as infeasible due to the counterintuitive nature of quantum theory. Even though quantum physics contains less information than classical physics, it is argued that the many strange and counterintuitive concepts that are involved in the former can hardly be accounted for in terms of classical physics. It should be

recognized however, that one is not obligated to account for the details of quantum theory. All of quantum mechanics (up to relativistic considerations) can be derived from the Heisenberg uncertainty principle, or from the logically equivalent Schrodinger formulation. The essential element is simply the non-zero value of the quantum of action.